# Localized spin waves at low temperatures in a Cobalt Carbide nanocomposite

Nirmal Roy[1], Arpita Sen[2], Prasenjit Sen[2], S. S. Banerjee[1,*]

*email: satyajit@iitk.ac.in

[1]Indian Institute of Technology Kanpur, Kanpur 208016, Uttar Pradesh India

[2]Department of Physics, Harish Chandra Research Institute, Jhunsi Allahabad 211019, UP.

**Abstract:** We study magnetic, transport and thermal properties of Cobalt carbide nanocomposite with a mixture of $Co_2C$ and $Co_3C$ phases in 1:1 ratio, with an average particle diameter of $40 \pm 15$ nm. We show that the behavior of the nanocomposite is completely different from that of either $Co_3C$ or $Co_2C$. We observed that with decreasing temperature the saturation magnetization $M_S(T)$ increases, however, below 100 K, there is a steep rise. A detail analysis shows the increase in $M_S(T)$ down to 100 K is explained via the surface spin freezing model. However, below 100 K the steep increase in $M_S(T)$ is explained by a finite size effect related to a confinement of spin waves within the nano particles. The measurement of heat capacity shows broad peak at 100 K along with presence of another anomaly at a lower temperature 43 K($=T_{ex}$). Resistance measurement in the nanocomposite shows metallic behavior at high $T$ with an unusual anomaly appearing at $T_{ex}$, which is near the $T$ regime where $M_S(T)$ begins to increase steeply. A measurement of the temperature gradients across the sample thickness indicates an abrupt change in thermal conductivity at $T_{ex}$ which suggests a phase transition at $T_{ex}$. Our results are explained in terms of a transformation from a magnetically coupled state with a continuous spectrum of spin waves into a magnetically decoupled state below 100 K with confined spin waves.

**INTRODUCTION**

Behavior of magnetism at nanoscales can be very different from that in the bulk form of the material, leading to exotic magnetic states. At low dimension due to enhanced surface to volume ratio, the properties are different from that in bulk. The predominance of surface atoms compared to bulk leads to a reduced co-ordination number, which in turns results in changes in the electronic band structure and also scenario's related to localization of electrons which may lead to enhanced moments [1,2]. For a small few atom cluster of ferromagnetic transition metals, the average magnetic moment increases with cluster size and applied magnetic field [1,2,3]. Ferromagnetic materials at nanoscales exhibit modified coercivity and anisotropy energy compared to bulk. Study of these magnetic nano-materials is important for their potential for applications in high density data storage, drug delivery and other industrial applications [4,5,6,7]. In particular nanomaterials of transition metal carbide systems (TMCs, M = Fe, Ni, and Co) have received attention in recent times because of their exotic electronic and magnetic properties coupled with wide potential for application as ferro fluids, catalysis, bioimaging and memory devices [8,9,10]. Amongst these TMC, cobalt carbide nanoparticles have received more attention due to their high coercivity and remanence at room temperature along with their cost effective synthesis routes, making them rivals to rare earth based permanent magnets. Synthesis protocol of cobalt carbide nanoparticle system has been studied extensively and have fairly good control over the size and shape of the synthesized nanoparticles [4,11,12,13]. Their ferromagnetic Curie temperature ($T_c$) is ~ 500 K and dissociation temperature is ~ 700 K [14]. Cobalt carbide has two well-known forms, viz., $Co_2C$ and $Co_3C$. Electronic band structure calculations of bulk $Co_2C$ system [15] showed it is paramagnetic and strongly metallic in character. Recent theoretical studies showed emergent magnetic behavior in few atom small clusters of $Co_2C$ and $Co_3C$ [16].

These studies show that both $Co_2C$ and $Co_3C$ phases are magnetic with magnetic moment of 0.99 $\mu_B$/atom and 1.67 $\mu_B$/atom respectively. A wide range of coercivity from 450 to 1200 Oe have been reported along with low saturation magnetization ~ 13 emu/g for $Co_2C$, while for pure $Co_3C$ coercivity ranging between 1.6 to 2.1 kOe and saturation magnetization of 55 emu/g have been reported [16]. Pure $Co_3C$ nano particles shows unusually large magnetic anisotropy ~ $7.5 \times 10^5$ J/m$^3$ with 8 nm sized ferromagnetic domain and a $T_c$ of 573 ± 2 K [17]. While these studies have all been in the high temperature range, there exist very few studies on nanocluster of $Co_2C$ & $Co_3C$ at low temperature. Furthermore systematic study of magnetic, thermal and electrical properties of such systems are few.

We explore the magnetic, transport and thermal characteristics of a nanocomposite of a transition metal carbide material, namely Cobalt carbide. The nanocomposite, synthesized using a standard one-pot polyol reduction process shows a nearly 1:1 ratio of homogeneously admixed $Co_2C$ and $Co_3C$ phases. SEM and TEM analysis reveals the composite consists of clusters with an average diameter of 40 ± 15 nm. The nanocomposite is ferromagnetic with a coercive field of about 400 Oe at room temperature (*T*). The temperature dependence of $M_S(T)$ shows two distinct regimes, viz., a regime I, below 100 K corresponding to magnon confinement regime within the nano composite, and another regime II above 100 K, associated with freezing of surface spin along the saturation magnetic field direction. The behavior of $M_S(T)$ in regime I correspond to the onset of confined spin waves with discrete energy spectrum, with a minimum energy gap between the states is estimated to be ~ 1.7 meV (about 20 K). We estimate these spin waves are confined within nanocluster regions of size about 6 nm in diameter (< 40 nm). Temperature dependent specific heat measurement confirms this transition to confined spin wave state via the presence of a cusp at 100 K. The same measurement also reveals a low temperature anomaly at $T_{ex}$. The resistance measurement reveals a strongly metallic behavior which decreases with lowering $T$ along with an anomaly appearing at $T_{ex}$. Our results are explained in terms of a transformation from a magnetically coupled state with a continuous spectrum of spin waves which transform below 100 K to a decoupled state with confined spin waves. We propose that exchange interaction which promotes the coupling is mediated by free electrons available in system at high $T$, which diminish with lowering of $T$.

## I. EXPERIMENTAL RESULT AND DISCUSSION

Cobalt acetate tetrahydrate (Co(CH$_3$CO$_2$)$_2 \cdot$4 H$_2$O), tetra (ethylene glycol) (TEG), sodium hydroxide (NaOH), polyvinylpyrrolidone (PVP) and absolute ethanol from Sigma-Aldrich Co, are used to synthesize the cobalt carbide nanoparticle clusters. Samples are synthesized via one-pot polyol reduction process [4]. First, 0.8 g of NaOH is dissolved in 20 mL TEG in a container by heating it to 100$^0$ C. Another 30 mL of TEG is taken into 100 mL European flask containing 2.5 mmol of Co(CH$_3$CO$_2$)$_2 \cdot$4 H$_2$O and 0.75 g of PVP. The mixture is then stirred for 20 min at room temperature by using magnetic stirrer. After that, NaOH is poured into the European flask, and the mixture is heated to 373 K for 30 min to remove water from solution. The solution is heated further to the boiling point of TEG (583 K) for one hour. The solution is allowed to cool down to room temperature naturally. After that, the solution is mixed with ethanol, and the precipitated nano particles are separated using a rare earth magnet from solution. The residual solution is drained, and ethanol is added to precipitate, and the process is repeated for several times. The extracted nanoparticles are dried in a vacuum oven and powders are compacted into pellets. A pellet of size (2.13×1.23×0.36) mm$^3$ is used for electrical transport measurements and another pellet of weight 4 mg is used for magnetization measurement. Powder X-ray diffraction (XRD) is performed on a pellet using Panalytical X-ray diffractometer with Cu Kα radiation ($\lambda$=1.5406 A°) at room temperature in the 2-theta range of $(30° - 90°)$. XRD pattern of the synthesized powders of nano particles is shown in Fig. 1(a). Rietveld refinement of this pattern is carried out using FullProf package. Rietveld refinement indicates the presence of Co$_3$C and Co$_2$C phases of cobalt carbide in almost equal proportion. Table I shows the comparison of the standard and refined values of *a*, *b*, *c* lattice parameter of both Co$_2$C and Co$_3$C. The comparison shows a slight deviation of the refined values w.r.t the standard lattice parameters. Figure 1(b) & (c) show SEM micrograph and TEM image indicating the particle size of 40±15 nm. Surface morphology in the Fig. 1(c) shows the synthesized particles aggregate to form clusters.

Table I. Comparison of lattice parameters of Co$_2$C & Co$_3$C phases.

| Lattice Parameter | Standard Value (Å) | | | Refined Value (Å) | | |
|---|---|---|---|---|---|---|
| | a | b | c | a | b | c |
| Co$_2$C | 4.3707 | 4.4465 | 2.8969 | 4.36778 | 4.44870 | 2.88908 |
| Co$_3$C | 5.07 | 6.7 | 4.53 | 5.11700 | 6.66900 | 4.49100 |

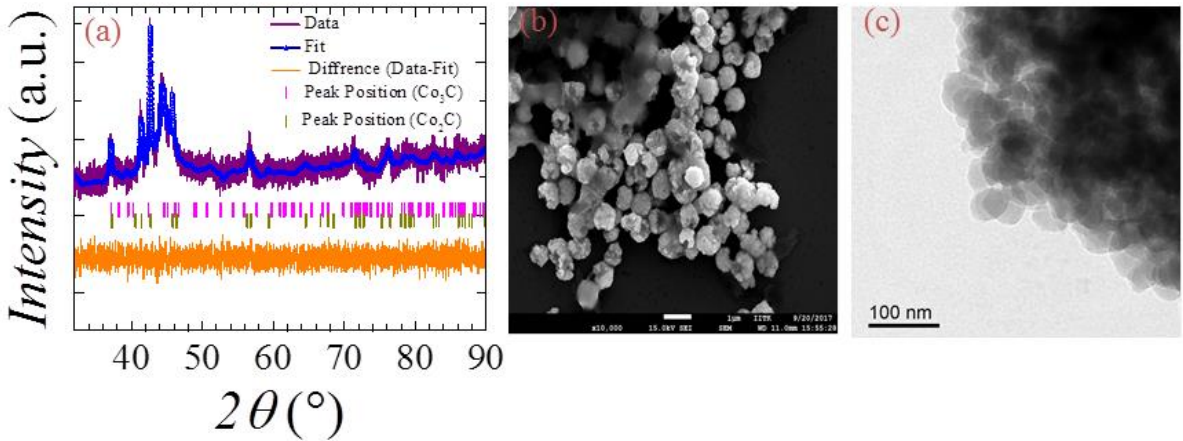

FIG. 1. (a) Powder X-Ray diffraction pattern (purple circle) of admixture of $Co_2C$ and $Co_3C$ nanocomposite system. The blue line, short vertical bars (Magenta and dark yellow colour correspond to $Co_3C$ and $Co_2C$ phases) and orange line represent the Rietveld refinement profiles, the fitted Bragg positions, and the residuals respectively. (b) SEM image of nanoparticles clusters and (c) TEM image of nanoparticles.

The XRD data analysis shows the absence of peaks matching with pure cobalt in the sample. The synthesis method of the nanocomposite is such that it is difficult to get completely pure phases.

Magnetization of the $Co_2C$-$Co_3C$ nanocomposite is studied as a function of magnetic field ($H$) at different temperatures ($T$) using the superconducting quantum interference device (SQUID) magnetometer (Cryogenics, UK). Figure 2(a) shows the ferromagnetic nature of the *M-H* hysteresis loops at different temperatures. Saturation magnetization ($M_S$) of the nano-composite at room temperature is about 36 emu/g which is much lower than bulk cobalt (~158 emu/g) suggesting the absence of pure cobalt in the sample [18]. X-ray diffraction analysis of the synthesized nanocomposite showed a homogeneous mixture of $Co_2C$ and $Co_3C$ nanoparticles in almost equal proportion. Using the magnetic moments of $Co_2C$, $Co_3C$ and $M_S(T)$ values, we reconfirm that $Co_2C$ & $Co_3C$ are admixed in a 55:45 % ratio (see supplementary information (SI 1)) in the nanocomposite sample. The observation of robust ferromagnetic hysteresis at room $T$ suggests no superparamagnetic features exist in the nano composites up to 300 K. Infact, the nano composite exhibits significant coercivity ($H_C$) of $\approx$ 400 Oe at 300 K. We also see a difference between field cooled and zero field cooled magnetization surviving up to room temperature indicating that the Blocking temperature ($T_B$) is greater than room $T$. The saturation magnetization ($M_S$) of the sample at different temperature is measured from the *M-H* hysteresis loops. Figure 2(b)&(c) shows $M_S$ increasing monotonically with decreasing $T$ however below 100 K the $M_S$ increases steeply. One possible source of increase in $M_S(T)$ is could be from paramagnetic contribution of $Co_2C$ phase at low $T$. In inset of the fig.2(a), we plot the dc susceptibility $\chi_{dc}(T)$ versus $\frac{1}{T}$. The shaded region in the figure is in the low $T$ regime (below 100 K). Clearly from below 100 K there is no regime where $\chi \propto \frac{1}{T}$ which represents absence of any dominant paramagnetic contribution appearing at low $T$. Another possible source of increase in $M_S(T)$ for ferromagnetic nano particles well below $T_C$ is because of the contribution of surface spin freezing along the saturation field direction [19,20]. This increase has been shown to be of the form $M_S(T) \propto \exp\left(-T/T_f\right)$. Such behavior of $M_S(T)$ has been reported earlier in oxide nanoparticles associated with a magnetically disordered surface spin population which "freezes" out to contribute to the magnetization only in large applied fields, and well below the superparamagnetic blocking temperature [19,20]. In fig. 2(c) the straight line above 100 K shows $M_S(T)$ obeying this form. In fig. 2(c), this spin freezing regime is denoted as regime II. However, we see in fig. 2(c) below 100 K, the increase in $M_S(T)$ is much more steeper than the behavior expected from the spin freezing model. Infact the data below 100 K fits to an exponential increase of the form $\exp(-E/k_BT)$, where $E$ is an energy scale.

Conventionally, the demagnetization of ferromagnetic materials well below $T_C$ is explained by the excitation of a continuum of spin waves in the material with energy $E_k$ which is proportional to the spin wave stiffness $D$, $(E_k = Dk^2)$ where $k$ is the continuum wave vector of spin wave excitation [21]. However, for nanomaterials due to finite size effects there is a quantization of spin wave spectrum into a discrete spectrum [21,22,23,24,25,26,27]. In nanoparticles the spin waves can get confined within the nanoparticles, leading to discrete spectrum of spin wave excitation energy. For example a cubic shaped particle with sides of length $d$, spin wave excitation energies can be roughly expressed as $E_n = Dk_n^2 = D\left(\dfrac{n\pi}{d}\right)^2$ [21,25]. The minimum diameter for confinement is about half the wavelength of a spin wave. The discrete spectrum of the quantized spin wave energy levels, results in activated thermal excitation of the spin waves with varying temperature. In this situation it has been shown that for confined spin waves modes, the $M_s(T)$ has a thermally activated decay form [21] of the type, viz.,

$$M_S(T) = M_S(0) - C\left[e^{\left(-\dfrac{E_1}{k_BT}\right)} + e^{\left(-\dfrac{E_2}{k_BT}\right)} + \ldots\ldots\right] \qquad (1)$$

where $E_1$, $E_2$ …. are the energy gaps in the discrete energy spectrum of the spin waves. Figure 2(b) shows the $M_S(T)$ below 100 K fits well with equation (1) with just a single exponent term. This is regime I in fig.2(b) Where $E_1$=1.75 meV, corresponding to a temperature scale of $T \sim 20$

K. Using $E_1 = Dk^2$ and $D$ is related to $B$ through $B = 2.3149 V_0 \left( \dfrac{k_B}{4\pi D} \right)^{3/2}$ where $V_0$ is atomic volume and $B$ is the Bloch'coefficient which govern the behaviour of saturation magnetization of the delocalised magnons (spin waves) [21]. The typical value of $B$ is of the order of $10^{-5}$ K$^{-3/2}$ [19,21]. Using this value of $B$ and $E = 1.75$ meV we determine the wave vector ($k$) of the spin wave ~ 0.48 nm$^{-1}$ from which the size of confinement within the nanoparticles ($d = \dfrac{\pi}{k}$) is estimated to be of the order of 6 nm. Thus at high T there is a continuum of spin waves modes, with lowering of $T$ the nanocomposite exhibits discrete spectrum of spin waves due to onset of confinement.

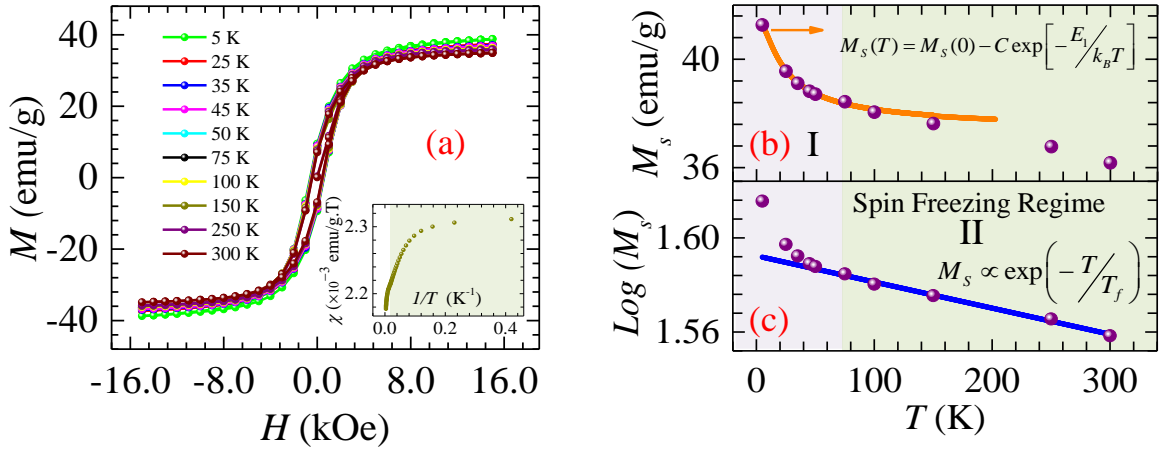

FIG 2. (a) Hysteresis loops of the nano composite system at different temperatures. Inset of the fig. 2(a) shows dc susceptibility ($\chi$) as function of $\dfrac{1}{T}$. (b) Saturation magnetization as a function of temperature, orange line is fitted with equation (1) below 100 K. (c) Temperature dependence of $M_S$ in log linear plot, blue line is straight line fitting up to 100 K.

To explore these issues further, we perform thermal and electrical transport measurements.

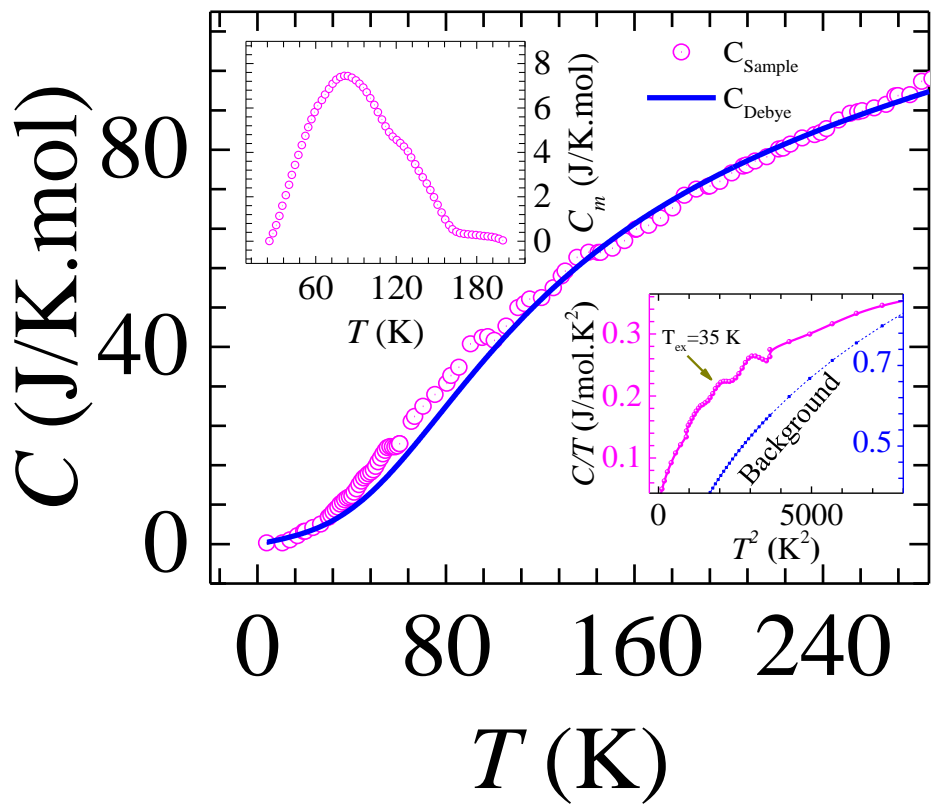

FIG. 3. Temperature dependence of specific heat (C) and solid line is the fitted with equation (2). Inset (upper left) shows the magnetic contribution ($C_m(T)$) of the heat capacity and Inset (lower right) represents the magnified viewed of $C/T$ vs. $T^2$ in 35 K-55 K regimes.

To probe the above features, we measured the temperature dependence of specific heat of this nanocomposite along with the background from the sample holder and system without the sample. The background measurement is performed by measuring the heat capacity without the sample. Note that for our background measurements, on the sample holder instead of the sample we place the amount of Epison used for gluing our sample to the holder. The specific heat of the sample ($C_{sample}$) shown in in fig.3 is obtained by subtracting the background specific heat from the total specific heat measured with the sample. We have analyzed $C_{sample}(T)$ to determine the magnetic contribution to specific heat.

$$C_{Sample} = C_{Debye}(T) + \gamma T + C_m(T) \tag{2}$$

where $C_{Debye}(T) = \dfrac{9nR}{\left(\theta_D/T\right)^3} \int_0^{\theta_D/T} \left(\dfrac{x^4 \exp(x)}{(\exp(x)-1)^2}\right) dx$, $\theta_D$ is Debye temperature, $n$ and $R$ are number density in a single formula unit of the materials and molar gas constant and $C_m(T)$ is the magnetic contribution to specific heat while $\gamma$ is the Sommerfeld's constant. The measured $C_p(T)$ data at low $T$ (< 10 K) and high $T$ (> 150 K) were fitted to $C_{Debye}(T) + \gamma T$, to obtain $\theta_D = 472$ K and $\gamma = 92 mJ/K^2 mol$. Using these fitting parameters the specific heat values were determined at all $T$ (blue curve in fig. 3). Finally, the magnetic contribution to specific heat was determined by $C_m(T) = C_{Sample} - \left[C_{Debye}(T) + \gamma T\right]$. Inset of fig.3 shows the behavior of $C_m(T)$. It is clear there is a significant increase in the magnetic contribution to specific heat as $T$ is lowered. Infact the $C_m(T)$ peaks close to 100 K which is the regime where $M_s(T)$ in fig. 2(b) had shown the beginning of exponential increase due to localization of spin waves. The peak in specific heat near 100 K we believe signals the onset of the scenario where a delocalized spin wave state at high $T$ enters a localized spin wave state at low $T$. It is interesting to note within 35 K-55 K the lower right of the inset of the Fig. 3, shows modulations in the C/T vs. $T^2$ curve of the sample (the average $T$ location of the modulations is marked a $T_{ex}$, cf. Fig. 3 inset). We shall show below the modulations in specific heat at $T_{ex}$ coincide with features in the electrical transport measurements. The smooth behaviour of the background in the inset confirms that the modulations in the C/T vs. $T^2$ near $T_{ex}$ in the low $T$ range arise from the sample. The modulation in the vicinity of $T_{ex}$ suggests transformations in decoupled state at low $T$. The break in slope of C/T vs. $T^2$ of the sample and the average relatively steeper fall of C/T vs. $T^2$ below $T_{ex}$ (inset of Fig. 3) suggests suppression of available density of states.

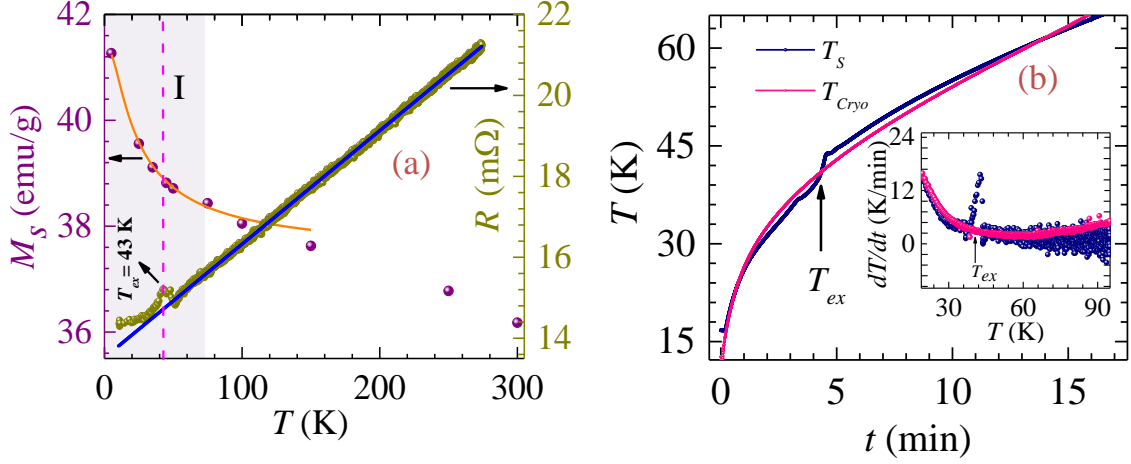

FIG. 4. (a) Temperature dependence of resistance of admixture nanocomposite system. The left side axis of Fig.4 (a) shows the exponential increase due to confinement of spin waves in $M_s(T)$ is in the same $T$ regime of modulation in $R(T)$ at $T_{ex}$ and blue line in the linear fit. (b) Variation of temperature of sample stage ($T_{Cryo}$) without sample and the sample ($T_S$) with time and inset is the corresponding heating rate with temperature.

We study the $Co_2C$ & $Co_3C$ nanocomposite using resistance verses temperature $R(T)$ and thermal measurements. Note that from the room temperature resistance ($R$) measurement, the average bulk resistivity of the nanocomposite material is determined to be $\rho \approx 10^{-6}$ $\Omega$m, which is nearly two orders of magnitude higher than that of pure cobalt. Our XRD analysis and also saturation magnetization value conclude the absence of any significant amount of pure Co in the nanocomposite. Hence we do not believe there are any percolating channels of unreacted Co in the nanocomposite leading to the features to be discussed, and neither can it explain all the results reported in this paper. Figure 4(a) shows the $R(T)$ behavior of the $Co_2C$ & $Co_3C$ nanocomposite pellet measured down to 10 K with a current of 20 mA. Note with decreasing $T$ the metallic character of the nanocomposite decreases rapidly. With lowering of $T$ along with the exponential increase in $M_S(T)$ due to confinement of spin waves we see a change in curvature of $R(T)$ at $T_{ex}$ = 43 K (where modulation was observed also in the specific heat data of the sample). In the supplementary information, SI 2 section, we show the presence of the modulation in $R(T)$ in the vicinity of $T_{ex}$ in another sample. Thus, the feature in $R(T)$ at $T_{ex}$ is seen in different samples at nearly, the same location with similar features. The figure shows at low $T$ (below $T_{ex}$) there is a significant difference between the actual $R(T)$ data and the extrapolated linear blue line from above 100 K, indicating presence of the excess resistance in the low $T$ state. Thus, the transport measurements show that at high $T$ there is a metallic response of the nanocomposite which decreases rapidly with decreasing $T$ suggesting a decrease in the density of states available for the free electrons at low $T$. Below $T_{ex}$ there seems to a transformation into a low conducting phase.

Next, we place a calibrated Cernox temperature sensor directly in thermal contact with the sample surface to measure its temperature ($T_S$) while the temperature of the cold finger of the cryostat $T_{Cryo}$ (in the absence of a sample) is monitored via another calibrated Cernox sensor (note $T_{Cryo}$ value when measured with the presence of a sample or in its absence shows no difference within experimental errors of ±10mK). The behavior of both $T_S$ and $T_{Cryo}$ is shown in Fig. 4(b) as a function of time, as the sample is warmed up towards 100 K. We observe that there is a sudden increase in the sample $T$ at $T_{ex}$ = 43 K (where we observed a modulation in specific heat and a feature in R(T)). The comparison of $\frac{dT}{dt}$ vs. $T$, both with and without the sample in the inset of Fig. 4(b) clearly shows the presence of a jump in $T_S$ at $T_{ex}$. Thus the above experiments suggest that below 100 K we believe is related to a phase transformation in the decoupled magnetic state of the system near $T_{ex}$.

The intriguing feature near $T_{ex}$ is explored in Fig. 5 which shows the $M_S(T)$ behavior while heating the sample from low $T$ and while cooling the sample from high $T$, with heating and cooling rates of 1.2 K/min. We observe the cooling and heating curves bifurcate above $T_{ex} = 43$ K, with the heating curve falling below the cooling curve. With heating from below $T_{ex}$, decoupled confined spin wave state seems to super heat to above $T_{ex}$ with an average lower magnetization which it had below $T_{ex}$. This suggests the possibility to supercool or superheat across $T_{ex}$. The above result along with this suggest the possibility of a first order like phase transition present in the nanocomposite below $T_{ex}$.

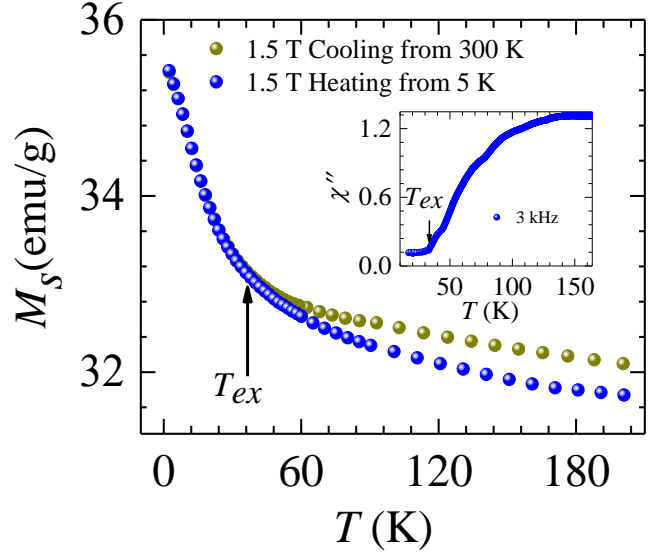

FIG. 5. Magnetization of admixture nanoparticle as a function of temperature during cooling (dark yellow) and heating (blue) cycle. Inset of Fig. 5 shows temperature dependence out of phase component of ac susceptibility ($\chi''$) at 3 kHz.

From our nanocomposite results, by using the size of the magnetized domain as equal to the size of the nanoparticle within which spin wave is confined, i.e., $d \sim 6$ nm and the typical reported value of $T_B$ for cobalt carbide is typically~ 578 K [17], we estimate the effective magnetic anisotropy constant for our nano composite is $K_{eff} = \frac{25 k_B T_B}{V}$, to be $1.76 \times 10^6$ J/m$^3$ (where $V$ is the domain volume $= \frac{4\pi}{3}\left(\frac{d}{2}\right)^3$). This magnetic anisotropy is much higher than pure cobalt and also much higher than that for pure Co$_3$C nanoparticle [7] Thus, it's clear that our results cannot be explained by response from only Co$_2$C or only Co$_3$C and the properties of the Co$_3$C-Co$_2$C mixture is very different from the individual components. Recall pure Co$_3$C phase behaves as a ferromagnetic while the Co$_2$C which is paramagnetic with a strongly metallic character. We propose at high $T$ presence of significant free charge carriers in the clusters with heterogeneous mixture of $Co_3C$-$Co_2C$ phase mediate RKKY like exchange interactions between $Co_3C$ molecules each of which possess large magnetic moments of 1.67$\mu_B$. Thus one has a strong magnetically coupled nanocomposite with deconfined spin waves at high *T*. Our *R*(*T*) measurements and specific heat measurements suggest decreasing metallicity and changes in the DOS of the free electrons with lowering of *T* below 100 K. Thus with fewer free electrons we believe the exchange interaction weakens clusters gets decoupled at low *T*. The transformation from coupled state to decoupled state at 100 K is a magnetic transformation associated with a bump in Specific heat. In decoupled state, we also observe the onset of the localization of spin waves below 100 K.

To confirm that at low *T* one has decoupled magnetized nanoclusters which couple at higher *T* (*T* > $T_{ex}$ = 43 K), we measured the out of phase component of ac susceptibility ($\chi''$) as a function of *T* for the nanocomposite sample. Figure 5. inset shows that as *T* falls below $T_{ex}$, the $\chi''(T)$ abruptly drops to negligible value compared to that above $T_{ex}$. Since, $\chi''$ is a measure of dissipation in the magnetic system, one will get significant value of $\chi''$ only if there is a sufficient magnetic correlation in the sample. At high *T*, as the clusters are collectively magnetically coupled, we observe a large saturated $\chi''(T)$ behavior above 100 K. Whereas below $T_{ex}$ as the clusters decouple, due to diminishing magnetic correlation amongst the clusters, the $\chi''(T)$ approaches zero. This feature provides an indirect confirmation that the nanocomposite is magnetically decoupled at low *T* and magnetic coupling is established at higher *T*. At the moment we are unsure what the low temperate phase below $T_{ex}$ could be. We speculate it could be a disordered glass configuration. In fact, like $T_B$, the spin glass transition temperature for this system we believe is well above room temperature.

# CONCLUSION

In conclusion we believe the $Co_3C$-$Co_2C$ nanocomposite exhibits features which are completely different from the individual constituents of the nanocomposite, viz., of only $Co_2C$ or $Co_3C$. The heterogeneous cluster with individual particles behaves like a strongly magnetically coupled entity. Such a state we show has collective excitations with long wavelength spin waves. However at low $T$ the nanocomposite magnetically decouples. Such a transformation at low $T$ triggers localization of small wavelength spin wave modes, causing opening of gap in the magnetic excitation spectrum leading to modified temperature decay of saturation magnetization. Such features of this cobalt carbide nanocomposite make it important for potential applications for magnetic memory storage and information transport.

**Supplementary material:**

Supplementary materials includes two sections.

Section SI 1 includes an estimate of ratio of Co3C to Co2C using the magnetic moments of $Co_2C$, $Co_3C$ and $M_S(T)$ values.

Section SI 2 shows the measurement of resistance anomaly in $R(T)$ at $T_{ex}$ (see fig.4(a)), reproduced in two different samples of the nanocomposite.

**Acknowledgements**

SS Banerjee would like to acknowledge funding support from IIT Kanpur and Department of Science and Technology- India, AMT and Imprint II programs.